\documentclass{ws-procs9x6}

\newcommand{\thp}{$\Theta^+$ }
\begin{document}

\title{Workshop Summary: Experiment}

\author{KENNETH HICKS
\footnote{\uppercase{S}upported in part by \uppercase{RCNP} 
(\uppercase{J}apan) and the \uppercase{NSF}/\uppercase{DOE} 
(\uppercase{USA}).}}

\address{Department of Physics and Astonomy \\
Athens, OH 45701, USA \\
E-mail: hicks@ohio.edu}

\maketitle

\abstracts{
A summary of experimental results from the Pentaquark 2004 
Workshop held at the SPring-8 facility in Japan is given. 
New results from the LEPS collaboration are highlighted, 
and older results are reviewed.  Non-observations are 
also discussed in light of theoretical estimates of 
possible $\Theta^+$ production mechanisms.  The problem 
of the narrow width and the parity of the $\Theta^+$ 
are explored and point to future experimental work 
that is needed.
}

\section{ Introduction }

This was an exciting workshop with many new results in the 
rapidly changing field of exotic baryons.  Both theoretical 
and experimental advances have been made, with new ideas
by the theorists to explain how a very narrow 
resonance can be constructed.  On the other hand, the 
experimental results are mixed, with some new positive 
evidence and some new null measurements, and little 
hope to clarify the questions of width and parity of the 
\thp within the next few years. Clearly, the existence of 
the pentaquark is an experimental issue, 
and it must be resolved before the physics 
community can take seriously the theoretical explanations. 
Here, I will focus on the experimental results 
and leave the theoretical summary to Carlson \cite{carlson}.

With both positive and null measurements of several 
possible pentaquark resonances, there are strong statements 
being made on both sides of the argument.  In fact, some 
might even say that it has become an emotionally-charged 
issue in hadronic physics.  It is important to realize 
that it takes time to do good experiments, and that nature 
sometimes has surprises in store for us.  For this reason, 
caution and patience are advised while we wait for progress. 
If we let science take its course, then in the end the truth 
will emerge.  

Since there is uncertainty in the existence 
of the pentaquark, the best we can do is ask whether there 
is good reason to be {\it optimistic} or {\it pessimistic} 
for the future of the pentaquark.  I will return to this 
question periodically.

\section{ Reasons to be Optimistic }

Over ten experiments have published evidence for the \thp pentaquark 
\cite{leps,diana,clas-d,clas-p,saphir,itep,hermes,zeus,na49,cosy,svd}.
While the number of measurements may be impressive, one must keep in 
mind that most of these are in the range of 4-6 standard deviations 
above a background that is difficult to quantify.  In fact, these 
statistical estimates {\it assume} a smooth background shape, and 
hence the statistical significance may be overestimated. Still, 
it seems unlikely that all of the results can be explained as a 
statistical fluctuation of the background, since so many different 
reactions (with different backgrounds) have been used in the analysis. 
Of course, such hand-waving arguments are not proof that the 
\thp exists, but the variety of reactions and the quality of 
the experiments does provide some encouragement that the \thp 
exists.

The HERMES and ZEUS experiments are well-known and respected (as 
are other groups that have published positive evidence).  However, 
these experiments have been criticized \cite{meson04} because they 
cannot determine the strangeness in the $K^0p$ final state.  In 
this workshop, the HERMES collaboration \cite{lorenzon} showed 
that the $K^0p$ peak in their results is {\it not} consistent with 
an interpretation as a $\Sigma^{*+}$ resonance.  They also showed 
that the peak to background ratio in their data can be enhanced by 
applying cuts that remove the known $K^*$ and $\Lambda^*$ resonances.
This strengthens their case that the peak is real, although more 
statistics are needed, which will occur in the next year. 
This is some reason for optimism.

An intriguing result was presented by the GRAAL collaboration
\cite{schaerf}
which did not search for the \thp but instead have evidence for a 
narrow $N^*$ resonance near 1680 MeV.  In their measurement, 
$\eta$-photoproduction on deuterium, they separate events that 
occurred on the neutron or the proton.  In a theoretical prediction 
by Polyakov and Rathke \cite{polyakov}, transitions from octet 
baryons to antidecuplet baryons are suppressed on the proton 
(due to an isospin factor) but allowed on the neutron.  In fact, 
this is what the GRAAL collaobration see, albeit with limited 
statistics.  If this claim can be confirmed by other experiments, 
then this narrow $N^*$ resonance fits better into the $\overline{10}$ 
symmetry group that includes the \thp as well.  But until the 
GRAAL result is confirmed, it is wise to resist the urge for optimism.

Perhaps the best evidence so far for the \thp was shown at this 
workshop by the LEPS collaboration \cite{leps-d}.  This analysis 
of the $\gamma d \to K^+ K^- X$, where $X$ is restricted to have 
the mass of deuterium, is an improvement over analysis shown at 
the MESON04 conference \cite{meson04}.  Additional cuts to remove 
coherent production and an energy-dependent $\phi$-meson exclusion 
were shown to enhance the $\Lambda(1520)$ peak in the $K^+$ missing 
mass spectrum (corrected for Fermi motion).  These same cuts also 
enhance the peak in the $K^-$ missing mass, where the \thp might 
be expected.  A new development is the use of event mixing to get 
the shape of the background.  Event mixing uses a $K^+$ from one 
event and a $K^-$ from a different event.  The missing mass of 
the mixed events is still required to be at the deuterium mass, 
which ensures energy conservation.  The advantage of using event 
mixing is: (1) the statistics can be increased because of the 
number of combinations, (2) the real angle and momentum distribution 
is used for the kaons, which is better than a phase-space Monte 
Carlo that is sometimes used to determine the detector acceptance, 
(3) correlations between the $K^+$ and $K^-$ are removed, so that 
the generated background goes smoothly under real peaks (which 
by definition have correlated $K^+K^-$ pairs).  The event mixing 
was shown \cite{leps-d} to work for the background under the 
$\Lambda(1520)$ and using the same procedure, the \thp peak also 
comes up clearly above the mixed-event background.  This  
quantitative analysis of the background shape  gives more 
credence to the \thp peak seen in the new LEPS deuterium data.

Finally, the best reason for optimism is that there are several 
new results on the horizon that have the potential to convincingly 
settle the question of whether the \thp exists or not.  The 
CLAS collaboration has taken new data on both a deuterium target 
\cite{tedeschi} and a proton target \cite{devita} with about 10 
times the statistics of earlier data, and expect to have results 
on several different \thp search topologies by early 2005. The 
COSY-TOF collaboration \cite{eyrich} will upgrade their detector 
and will take more data in 2005, thus increasing their statistics 
by (perhaps) a factor of five.  As already mentioned, the HERMES 
collaboration will double their statistics on a deuterium target 
soon, which can substantially help their \thp search.  With these 
new results on the horizon, 2005 will be an exciting year.

\section{ Reasons to be Pessimistic }

There are a number of experiments that give null results in a 
search for the \thp pentaquark.  At the time of this conference, 
only 2 were published \cite{bes,hera-b} and a number had been 
presented at an earlier conference \cite{indiana}.  At this 
meeting, there were presentations of pentaquark searches by the 
BaBar \cite{babar}, Belle \cite{belle} and Fermilab E690 \cite{e690} 
collaborations.  Naively, one would expect to see the \thp (and 
perhaps the $\Xi^{--}$) in these experiments, and the null results 
imply that either: (1) the pentaquark does not exist, or (2) the 
production mechanism of pentaquarks differs from that of 3-quark 
baryon resonances.  In any case, these results are good reason 
for one to be pessimistic (or at the very least very cautious) 
about the existence of pentaquarks.

The BaBar results \cite{babar} have high statistics and reasonable 
signals for the established $N^*$ and $Y^*$ resonances.  Here, a 
baryon-antibaryon pair is created from $e^+e^-$ collisions at 
$\sqrt{s}=10.58$ GeV. The $\Lambda (1520)$ resonance is seen clearly 
in the $pK^-$ invariant mass, but no structure is seen in the $pK^0$ 
mass spectrum in the region of the \thp mass.  Based on systematics 
of baryon production rates as a function of baryon mass, one can 
estimate the number of \thp baryons that should have been produced. 
However, this assumes that pentaquark production (involving 5 quarks 
and 5 antiquarks) follows the same systematics as 3-quark baryons. 
{\it Due to the uncertainty in the production mechanism, theoretical 
calculations are needed to understand the true significance of these 
null results.}

The Belle experiment \cite{belle} took a different approach.  They 
used secondary scattering of mesons (from $e^+e^-$ collisions) in 
their silicon vertex detector to produce known $Y^*$ resonances. 
If the \thp exists, it could be produced with a $K^+$ beam of the 
right energy.  Unfortunately, the hadrons incident on the silicon 
target have unknown identity and unknown energy.  Only a small 
fraction of these data could result in production of the \thp and 
detected by its decay into the $pK^0$ channel.  With the high 
resolution of Belle, even a small signal (with a narrow width) 
might be visible, but none was seen. Again, {\it we need better 
calculations of the expected number of counts (based 
on Belle's spectrum of hadrons incident on silicon)  before we 
can interpret their null result.}

The E690 experiment \cite{e690} uses protons of about 800 GeV in 
peripheral collisions with target protons.  By putting cuts on 
the missing transverse momentum and the longitudinal energy, 
exclusive reactions can be measured.  In the $pK^-$ mass spectrum, 
about 5000 $\Lambda (1520)$ events are seen, but no structure 
is seen in the $pK^0$ mass spectrum.  Because of the exclusive 
reaction, the strangeness of the $pK^0$ system is known and so 
this spectrum is not contaminated with $\Sigma^*$ resonances, 
such as the $\Sigma (1660)$.  Hence, this is a significant null 
result, and suggests one should be pessimistic about the \thp existence. 

One interesting development of this workshop was a calculation 
presented by Titov \cite{titov}, using quark constituent counting 
rules to estimate the ratio of \thp to $\Lambda (1520)$ production 
in fragmentation reactions.  Fragmentation functions are well 
established \cite{close} and have been used for years to describe 
the distribution of hadrons from high-energy collisions, based 
on the number of constituent partons in the projectile and target. 
Using this model, Titov shows that production of the $\Theta^+$ 
is suppressed relative to the $\Lambda$(1520) resonance by about 
3 orders of magnitude for experiments such as E690 and BaBar. 
Of course, the simple model used for this estimate may not be 
a good approximation for all kinematics, but it is consistent 
with the null experimental results at high energies.
The optimist would argue that we could have expected null 
results from fragmentation reactions in high-energy experiments.

It is easy to see that there is reason for pessimism, but the 
evidence is not entirely convincing. It is difficult to 
know how many \thp events should have been seen in the 
high-energy experiments with null results.  In fact, there is 
even a reasonable explanation from Titov, using fragmentation 
functions, for these null results.  The case for ``killing" 
the \thp has not been made.

\section{ The Problem of the Width }

Perhaps the most disturbing fact of the \thp evidence is 
that its width appears to be very narrow. Direct evidence 
\cite{diana,hermes,zeus} limits the width to be less 
than about 10 MeV.  Indirect evidence, based on analysis 
of KN scattering data \cite{nussinov,cahn,arndt,jeulich,gibbs}, 
estimates the width at a few MeV or less.  Such a narrow 
width for a resonance 100 MeV above its strong decay threshold 
would be unprecedented. 

Coupled with the narrow width problem is the 
question of parity.  The spin of the lowest-lying \thp 
is expected to be $J=1/2$ with either negative (S-wave) or 
positive (P-wave) parity. A narrow width from an S-wave 
resonance makes no sense \cite{jw} whereas a P-wave 
would allow a centrifugal barrier making a narrow width 
at least possible \cite{jw,kl}.  It seems logical that 
if the \thp width is narrow, its parity must be positive. 
This idea was beautifully presented by Hosaka \cite{hosaka}.

What do lattice QCD calculations say about the parity?  
Several lattice results were presented at this conference, 
and except for one result \cite{chiu}, only the negative 
parity projection gives a result consistent with the \thp 
\cite{sasaki,lee}.
So we have an apparent contradiction between the parity 
deduced from quark models (above) and the parity deduced from 
(most) lattice calculations.  One obvious resolution to this 
dilemma is to conclude that the \thp does not exist. However, 
we must realize that the lattice calculations for exotic 
baryon resonances should be regarded as exploratory \cite{sasaki}. 
Extrapolating to the chiral limit from the heavy quarks 
used in lattice calculations must be done properly \cite{chiu,lee} 
and furthermore, all lattice calculations are done in the 
quenched approximation.  Hence we should be cautious about 
parity statements based on current lattice results.

Even if the \thp exists with positive parity, a width as 
narrow as 1 MeV is theoretically difficult to understand 
\cite{prasz}.  However, several new theoretical ideas 
were presented showing that such a narrow width is consistent 
with theory.  Using a two-state model, Lipkin showed \cite{lipkin} 
that the mass eigenstates of two pentaquarks ($e.g.$, mixtures 
of the Jaffe-Wilczek model and the diquark-triquark model) 
can mix, resulting in one coupling strongly to KN decay 
(with a wide width) and one decoupling (with a narrow width). 
Another approach, this time with mixing between the octet 
and the $\overline{10}$, shown by Praszalowicz \cite{prasz}, 
can suppress the width by a correction factor that depends 
on the value of the pion-nucleon $\Sigma$ term.  From a 
completely different angle, using the QCD string model, 
Suganuma {\it et al.} showed \cite{suganuma} that the 
pentaquark does not just ``fall apart" as predicted by the 
quark model, but must overcome a sizeable potential barrier 
to decay into a KN final state.  This results in a very 
narrow width for the \thp in their model.  In all, it is 
interesting that a narrow width of 1 MeV can be accomodated 
in the quark model, the chiral soliton model and the QCD 
string model.

Clearly, experimental information is needed before one can 
test the various ideas about the \thp width.  Proposals at 
KEK \cite{imai} and Jefferson Lab \cite{bogdan} for high 
resolution spectrometer experiments are being considered. 
Other facilities already mentioned (COSY-TOF, HERMES, ZEUS) 
will gather more statistics, which should enable a 
better determination of the \thp width.  In addition to 
width measurements, we need to know the \thp parity.  This 
will likely be done at COSY-TOF \cite{eyrich} using polarized 
target and polarized beam, which has a clear theoretical 
interpretation as shown by Hanhart \cite{hanhart}. If the 
\thp exists, then we have the experimental tools to learn 
about its width and parity.

\section{ Summary }

This workshop was filled with exciting new developments, both 
experimental and theoretical.  The LEPS collaboration showed 
new data, this time using a deuterium target, which appears 
to confirm the existence of the \thp although the results are 
still preliminary. A possible narrow $N^*$ state was seen in 
$\eta$ photoproduction from the neutron at GRAAL, but not from 
the proton, in agreement with theoretical predictions. While 
no new data were shown for the \thp width or parity, 
several theoretical models showed that a narrow width for the 
\thp is not unreasonable.  

The null results from high-energy experiments are worrisome, 
but theoretical estimates from fragmentation functions suggest 
that \thp production is very suppressed relative 
to baryon resonances like the $\Lambda$(1520).  If so, then 
this is not negative evidence for the \thp but just non-observation. 
Of course, the proof of \thp existence must be convincing in the 
medium energy experiments, with high-statistics data, before 
one can believe it is suppressed in high-energy data. 

So should we be optimistic or pessimistic about the existence of 
pentaquarks?  At the present time, there is no clear choice. 
However, new data will be available soon that will clarify 
the situation.  If the \thp exists, then we have a rich new 
spectroscopy to explore.

\section*{Acknowledgments}
I am grateful for the support I received from RCNP (Osaka University) 
and Kyoto University during my sabbatical in Japan.  It has been 
a great pleasure to work closely with Takashi Nakano and Atsushi 
Hosaka, along with the members of their groups, during my stay. 
I congratulate Hosaka-san for the excellent organization of the 
Pentaquark 2004 workshop.

\newcommand{\etal}{{\it et al.}, }

\end{document}